\documentclass[]{article}

\usepackage[paperwidth=8.5in, paperheight=11in, margin=1in]{geometry}
\usepackage{graphicx}
\usepackage[]{todonotes}
\usepackage{changes}
\usepackage[font=footnotesize,labelfont=bf,singlelinecheck=false,labelsep=period, skip=2pt]{caption}

\usepackage[hyphens]{url}            
\usepackage[breaklinks=true]{hyperref}
\usepackage{lineno}

\usepackage{subfigure}

\usepackage[backend=bibtex8,style=nature,citestyle=numeric-comp,sorting=none,maxbibnames=30,minbibnames=30,autocite=superscript,giveninits=true,terseinits=true]{biblatex}
\title{Organizing genome engineering for the gigabase scale}
\author{
Bryan A. Bartley\\
Raytheon BBN Technologies\\
\texttt{bryan.a.bartley@raytheon.com}\\[1em]
Jacob Beal\\
Raytheon BBN Technologies\\
\texttt{jakebeal@ieee.org}\\[1em]
Jonathan R. Karr\\
Icahn School of Medicine at Mount Sinai\\
\texttt{karr@mssm.edu}\\[1em]
Elizabeth A. Strychalski\\
National Institute of Standards and Technology\\
\texttt{elizabeth.strychalski@nist.gov}
}

\begin{document}
\maketitle

\begin{abstract}
Engineering the entire genome of an organism enables large-scale changes in organization, function, and external interactions, with significant implications for industry, medicine, and the environment. 
Improvements to DNA synthesis and organism engineering are already enabling substantial changes to organisms with megabase genomes, such as {\it Escherichia coli} and {\it Saccharomyces cerevisiae}. 
Simultaneously, recent advances in genome-scale modeling are increasingly informing the design of metabolic networks.  
However, major challenges remain for integrating these and other relevant technologies into workflows that can scale to the engineering of gigabase genomes.

In particular, we find that a major under-recognized challenge is coordinating the flow of models, designs, constructs, and measurements across the large teams and complex technological systems that will likely be required for gigabase genome engineering. 
We recommend that the community address these challenges by 
1) adopting and extending existing standards and technologies for representing and exchanging information at the gigabase genomic scale,
2) developing new technologies to address major open questions around data curation and quality control, 
3) conducting fundamental research on the integration of modeling and design at the genomic scale, and
4) developing new legal and contractual infrastructure to better enable collaboration across multiple institutions.
\end{abstract}

\section{From Engineering Genes to Engineering  Genomes}

\begin{figure}[t]
\centering
\subfigure[Exponential growth of engineered genome size]{\includegraphics[width=0.4\textwidth]{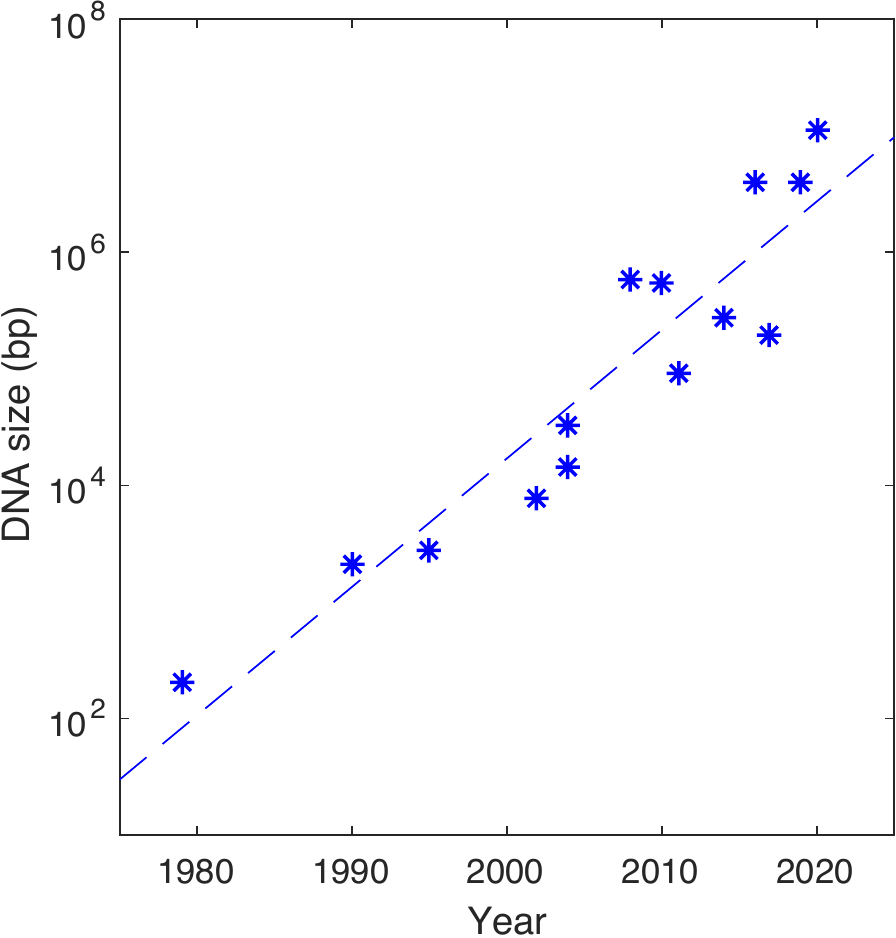}\label{fig:bp}}
\subfigure[Exponential growth of collaboration size]{\includegraphics[width=0.4\textwidth]{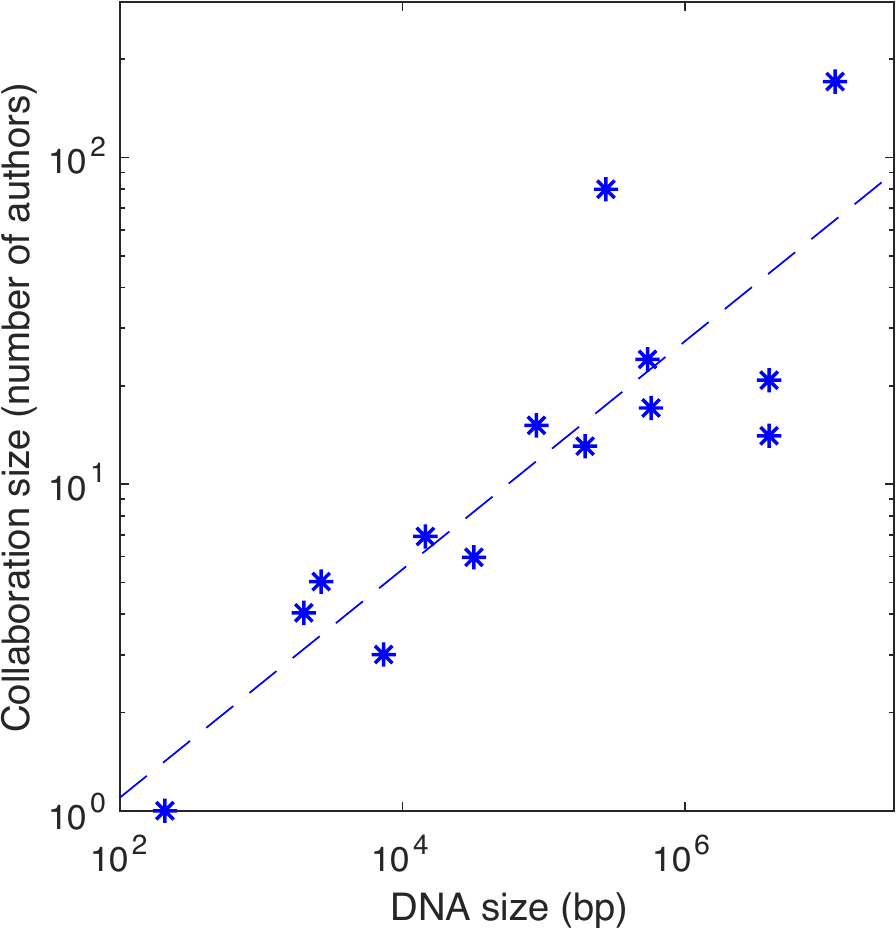}\label{fig:complexity}}
\caption{As capabilities for genome engineering advance rapidly, the size of teams involved in each genome engineering project also increase. 
(\textbf{a}) From 1980 to present, the size of the largest engineered genomes has grown exponentially, doubling approximately every three years. Extrapolating this trend projects gigabase engineering becoming feasible by 2050. 
(\textbf{b}) The sizes of the teams needed to produce these genomes has also grown exponentially, scaling with the cube root of genome size and suggesting that teams on the order of 500 investigators will be needed to engineer gigabase genomes. 
Data for this figure is provided in Supplementary Data 1.}
\label{fig:scale}
\end{figure}

Engineering the entire genome of an organism will allow large-scale changes in organization, function, and environmental interactions, with major implications for applications of biotechnology broadly~\cite{boeke2016genome}.
The past several decades have seen remarkable progress in our capability to create DNA and modify genomes~\cite{CarlsonCurve, hughes2017synthetic, chari2017beyond}.
Since Khorana created the first synthetic gene forty years ago~\cite{khorana1979total}, 
our capability to construct DNA sequences has doubled approximately every three years (Figure~\ref{fig:scale}A), progressing from plasmids in the early 1990's~\cite{mandecki1990totally, stemmer1995single}, 
viruses in the early 2000's~\cite{cello2002chemical}, and gene clusters in the mid-2000's~\cite{tian2004accurate, kodumal2004total}, to the first bacterial chromosome in 2008~\cite{gibson2008complete, gibson2010creation}.
More recently, several groups have demonstrated the feasibility of total synthesis of the 4\,Mb genomes of {\it Escherichia coli}~\cite{ostrov2016design, fredens2019total} and {\it Salmonella typhimurium}~\cite{lau2017large}, and 
the Sc 2.0 project~\cite{dymond2011synthetic, annaluru2014total} 
has nearly completed re-engineering the 11.4 Mb genome of {\it Saccharomyces cerevesiae}~\cite{richardson2017design}. 
Projecting further to the gigabase scale, in 2016 leaders from academia and industry formed the Genome Project-Write~\cite{boeke2016genome} consortium to roadmap new technical approaches and ethical frameworks for engineering the genomes of higher-order organisms. Since then, members of the consortium have proposed projects to develop human cell lines with engineered genomes, including a virus-resistant, ultra-safe cell line for pharmaceutical production~\cite{gpwrite2018}.

Moving to the gigabase scale poses major technological and scientific challenges, with scaling up DNA synthesis and/or editing arguably foremost among these.
These challenges related to synthesis and editing have been discussed extensively, including a recent summary in~\cite{TechnicalWorkingGroupInRevision}.
However, the challenges of managing the complex workflows and large teams needed for genome engineering not previously been analyzed.
Figure~\ref{fig:scale}B shows that the number of investigators needed to engineer a genome has also risen markedly with the size of the genome. If both of these trends continue, then the capability to engineer gigabase eukaryotic genomes would be projected to be realized in approximately 2050, with each such genome required a team of around 500 investigators.
Thus, engineering gigabase genomes will likely require new approaches to coordinate complex workflows and large, interdisciplinary teams.

Accordingly, we have examined the emerging design-build-test-learn workflow for genome engineering and identified potential bottlenecks and associated solutions, as well as areas where additional research will likely be needed.
In Section~\ref{s:workflows}, we discuss the emerging, design-build-test-learn workflow for organism engineering and identify key points to integrate workflow steps across processes and/or organizations.
We then discuss each integration point in detail in Section~\ref{state_of_the_art}, identifying gaps that must be addressed to support gigabase engineering.
Finally, in Section~\ref{recommendations}, we summarize and recommend priorities to advance genome engineering.

\section{Toward Workflows for Gigabase Engineering}\label{s:workflows}

Gigabase engineering will likely require large-scale integration of efforts for the division, distribution, and coordination of labor and materials. 
Both genome engineering and smaller-scale organism engineering workflows have often been abstracted and organized in terms of design-build-test-learn cycles~\cite{appleton2017needs, carbonell2018automated, hughes2017synthetic, gill2016synthesis, poust2014narrowing, pouvreau2018plant, hutchison2016design, richardson2017design}.
This iterative approach is helpful given the complexity and uncertainties in engineering biology, 
progressing in incremental steps,
implementing genetic modifications in stages, 
and adjusting designs based on information learned from testing prototypes and partial implementations.

\begin{figure}[t]
\centering
\includegraphics[width=0.9\textwidth]{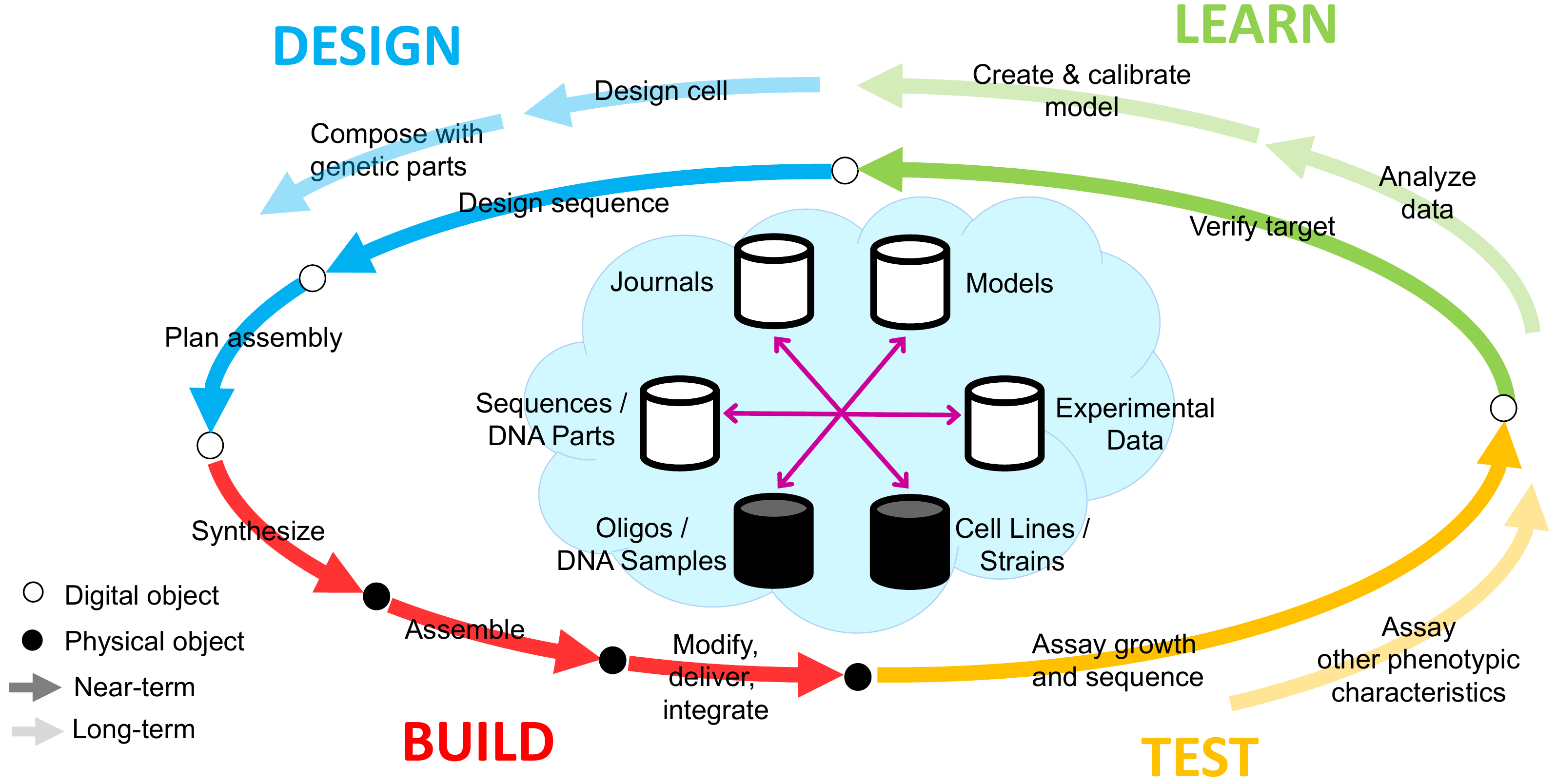}
\caption{The emerging design-build-test-learn workflow for genome engineering is shown schematically with current (solid arrows) and predicted (transparent arrows) tasks, interfaces (circles), and digital (white cylinders) and physical (black cylinders) repositories.}
\label{fig:workflow}
\end{figure}

Figure~\ref{fig:workflow} shows the emerging design-build-test-learn workflow for genome engineering.
The first step in this workflow is to \textit{design} a target sequence using models and design heuristics. The second step is to \textit{build} these sequences, producing and expressing DNA constructs. In the third step, the molecular and phenotypic characteristics and consequences of these sequences are \textit{tested}. Finally, researchers and engineers improve models, design heuristics, and processes by \textit{learning} from the data generated by testing, focusing on discrepancies between the predicted and observed results.

The primary loop in Figure~\ref{fig:workflow} indicates the workflow necessary for current genome engineering projects, which have largely focused on ``top-down'' approaches to recode and refactor genomes, such as reducing genomes to essential sequences.
As genome writing technology matures, the focus of genome engineering workflows may shift to more fully include the tasks in the outer loop of Figure~\ref{fig:workflow}.
For example, one ultimate aim of synthetic biology is to design novel genomes that encode new phenotypes from the ``bottom-up'' by assembling organisms from modular parts \cite{purnick2009second}.
Parts-based synthetic biology is already being used to engineer novel metabolic pathways for commercial production of high-value chemical products, but engineering on the genomic scale will be staggeringly more complex.
In this case, sharing and reusing intermediate, mid-scale DNA constructs that constitute biological parts and devices, as well as large-scale DNA constructs more akin to synthetic chromosomes, will likely become increasingly important. 

At each stage of the workflow, reagents, biological materials, and/or information must be transferred through an interface from one set of processes to another, often run by investigators with complementary areas of expertise and located in different groups or institutions. These products are both physical and digital in nature. 
For example, design steps should produce sequences described in a digital format, such as the Synthetic Biology Open Langauge (SBOL)~\cite{galdzicki2014synthetic,roehner2016sharing}, which then serves as an input for experimentalists to build those sequences. 
Build steps produce DNA constructs and cell lines to be tested. 
Test steps produce data to be learned from. Learn steps produce data-driven and mechanistic models, ideally expressed in a digital format. Materials and data at all stages may also reside in sample and data repositories. Genome engineers will need interfaces to these repositories to efficiently browse, access, alter, and transfer materials, as well as to cross-reference information across the stages of a given workflow. 

Beyond the technical challenges for each stage of the workflow, genome engineering must also contend with a number of cross-cutting issues to facilitate close coordination across many organizations. When technical information or materials are transferred, their recipients will need to know the associated contractual and legal obligations, such as information about intellectual property and licensing, safety, and any privacy or personally identifiable information (PII) concerns. Additionally, issues of cybersecurity, biosecurity, and/or biosafety may pertain at each interface.

For gigabase engineering, every one of the interfaces between tasks, repositories, and organizations will need to be optimized to mitigate bottlenecks to progress. 
For projects of this scale, ad hoc, human-centric, and bespoke interfaces will be impractical. 
Instead, every interface will require representations conducive to machine reasoning and automation, and appropriate accompanying cyber-infrastructure and tooling. Fortunately, most of the design-build-test-learn workflow is not unique to gigabase engineering, 
such that prior work in smaller-scale organism engineering provides a solid basis for 
paths forward from the state of the art towards realizing effective and routine engineering of large genomes.

\section{Identifying And Closing Gaps At The State Of The Art}\label{state_of_the_art}

In this section, we discuss the integration challenges identified in the previous section, 
reviewing the state of the art in technologies and standards with respect to the emerging needs of gigabase genome engineering. 
Instead of focusing on specific evolving protocols and methods, which are likely to advance rapidly, we consider the information that must be communicated to enable protocols or methods to be composed into a comprehensive workflow. 
Through this analysis, we identify critical gaps and opportunities, where additional technologies and standards would facilitate workflows that are able to effectively deliver gigabase engineered genomes.
Table~\ref{t:stoplight} summarizes these recommendations.

\begin{table}[h]
\centering
\includegraphics[width=0.9\textwidth]{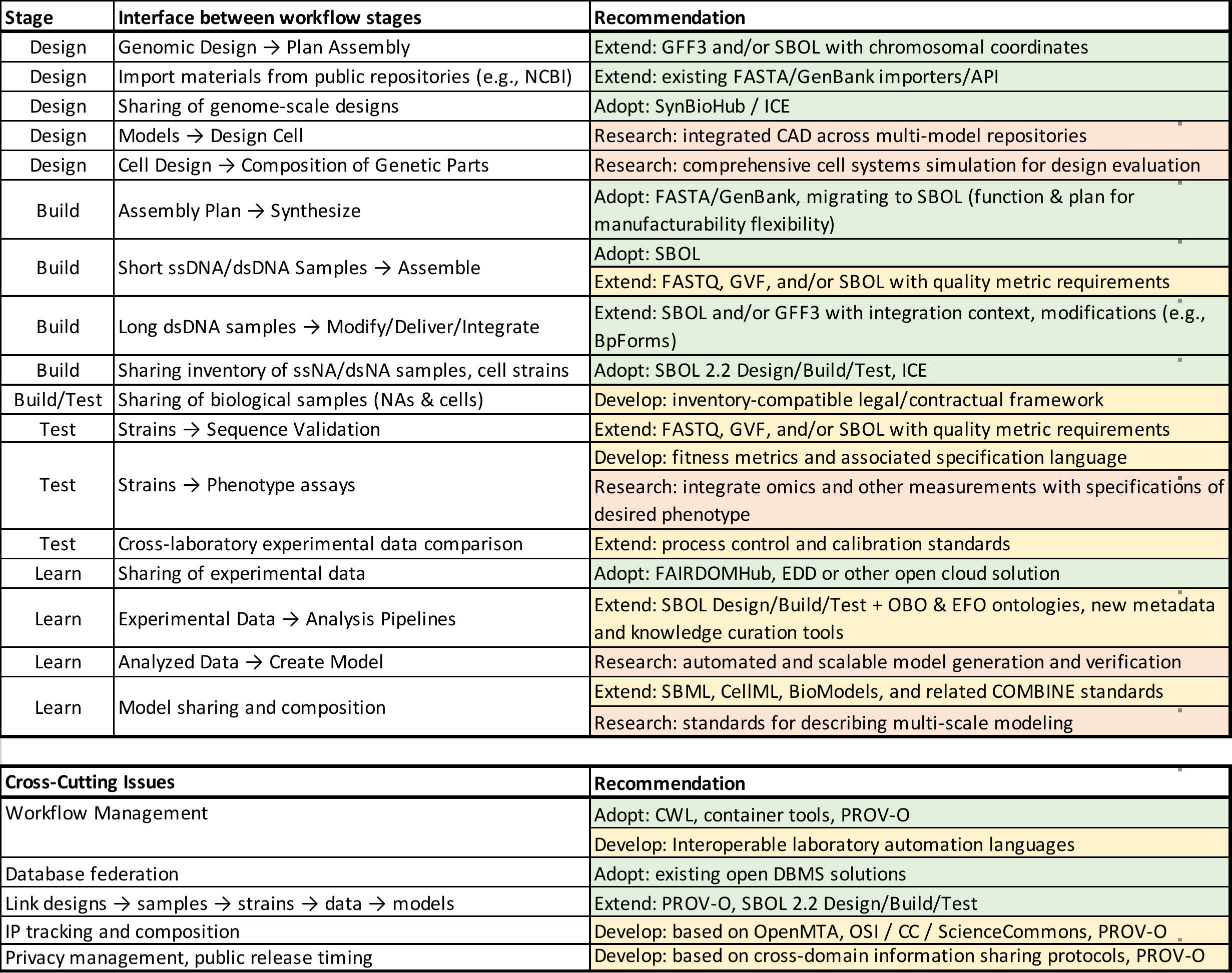}
\caption{We assess the state of the art of gigabase-scale genome engineering.
Recommendations are categorized into adopting or extending existing technologies for near-term solutions, 
developing new technologies for mid-term solutions, and the need for additional fundamental research for longer-term solutions.
Color indicates technology readiness level:
green indicates needs that can be fulfilled by adopting or extending relatively mature existing methods, 
yellow indicates potential solutions from extensions of the state of the art, 
and red indicates areas where more fundamental research is required.}
\label{t:stoplight}
\end{table}

\subsection{Rationally refactoring and designing gigabase genomes}

The first step of the design-build-test-learn workflow is to design a synthetic genome. Currently, most synthetic genomes are designed ad hoc using a combination of biological knowledge, engineering intuition, and informal design heuristics. This approach has enabled projects, such as reducing \textit{Mycoplasma} genomes by eliminating non-essential elements \cite{hutchison2016design}, 
reordering genes into functional groups \cite{annaluru2014total}, and inserting small metabolic pathways \cite{steen2008metabolic}. 
However, this approach can be inefficient, slow, and expensive. In contrast, most other engineering fields, such as aerospace engineering, use computer-aided design (CAD) powered by mechanistic models to reliably design complex systems, such as commercial jets and spacecraft. Ultimately, CAD powered by predictive models will likely be needed to reliably design novel genomes that encode novel functions \cite{sun2005bio}. In turn, this will require comprehensive and detailed models that can help design of entire genomes.
Substantial fundamental research is required to enable such model-driven design at the genome scale.

As genomes cannot currently be designed \emph{de novo}, for the foreseeable future, genome design will likely involve modifying the sequences of existing organisms. Annotated genome sequences are available for diverse organisms. Public archives, such as the National Center for Biotechnology Information, the EMBL European Bioinformatics Institute, and the DNA Data Bank of Japan, which collectively constitute the International Nucleotide Sequence Database Collaboration (INSDC)~\cite{cochrane2015international}, presently contain on the order of 10\textsuperscript{5} bacterial genomes and hundreds of eukaryotic genomes~\cite{cunningham2018ensembl, kersey2015ensembl, o2015reference, haft2017refseq, mashima2015dna}. 

Genome design and debugging can be further facilitated by supplementing genome sequences with additional functional annotations. For example, INSDC reference genomes primarily annotate all known transcriptionally active parts of the genome, but the engineer will also need to consider tissue-specific expression patterns.  Other important sequence features include intergenic elements, including regulatory elements, such as enhancers and silencers, structural elements, replication origins, and clinically-significant sites of DNA recombination and instability. Much of this knowledge that can help the engineer is unfortunately distributed among many independent resources. Services such as NCBI Genome Viewer~\cite{gdv}, WebGestalt~\cite{liao2019webgestalt} and DAVID~\cite{huang2009systematic} provide programmatic or web interfaces for integrating annotations. An ideal platform for genome design would integrate such resources with sequence editing and design tools.

The quality of annotations on a genome is another key consideration in the design process. Genomes can have significant differences in annotations depending on the toolchains that generate them, and these differences will likely result in trade-offs impacting engineering decisions. For example, the human reference genomes generated by the RefSeq and GENCODE projects have notable differences~\cite{frankish2015comparison, mccarthy2014choice}. 
GENCODE annotations have been shown to cover more coding regions and alternative splice regions and consequently are more likely to flag a loss-of-function when a sequence variant occurs~\cite{frankish2015comparison}. The sensitivity and specificity of annotation tools are thus important factors in sequence design and debugging. A genome engineering effort would therefore benefit from adoption of a standard convention for labeling annotations with estimates of confidence and reliability, such as the RefSeq database does with the Evidence and Conclusion Ontology~\cite{chibucos2017evidence}. 

Collaborative design of genomes design at large scale will require an appropriate choice of data formats and schema. Many commonly used sequence formats, such as GenBank and EMBL, are formatted for human readability, but have significant limitations with respect to managing large genomic designs. Transferring genomic data over a network in these formats can be a significant bottleneck, due to their monolithic treatment of sequences. Moreover, these formats are not easy to merge or harmonize across multiple concurrent users.  Fixed formats like GenBank and GFF do not facilitate integration of annotations from diverse resources, while open semantic formats, such as SBOL, more easily support such extensions.

For communicating a genomic design, however, there are already two good options to adopt and extend. 
One is the Generic Feature Format Version 3 (GFF), which allows hierarchical organization of sequence descriptions (e.g., genes may organized into clusters, and clusters into chromosomes) and makes use of the Sequence Ontology~\cite{SequenceOntology} for coherent sequence annotations.
GFF has already been used in the Sc~2.0 genome engineering project~\cite{richardson2017design}.
The other is the Synthetic Biology Open Language (SBOL)~\cite{cox2018synthetic}. 
Like GFF, SBOL supports hierarchical description with standardized vocabularies.
SBOL, however, also allows representation of abstract architectures in which sequences are not yet fully specified (e.g., asserting that a certain set of genes will be used, but not yet the particular variants or their arrangement), which is useful both for interchange during the design process and also for representing sequence variants and combinatorial libraries.
SBOL can also describe other molecular components and interactions in a cell (e.g., proteins, metabolites, regulatory interactions), allowing specification of phenotype and its linkage to genotype.
This also allows designs to be organized in terms of functional relations (e.g., metabolic pathways, cellular systems), rather than simply genomic proximity.
Finally, the SBOL standard is interoperable with SBML, allowing bidirectional conversion between designs and simulation-ready models.
For both GFF and SBOL, however, it would be useful to have a richer language for specifying the position of sequences in a chromosome: the current practice of using sequence index is fragile to unrelated distant changes impacting the alignment of design fragments.

To date, genome engineering projects have minimally leveraged tooling that integrates sequence-design and network-scale modeling.
Models will become increasingly important as the target genomes increase in scale or diverge from native architecture. For example, models could help predict growth phenotypes, including non-viable states, that may result from engineering sequence elements. 
Network-scale models, such as genome-scale metabolic models~\cite{swainston2016recon, monk2017iml1515} and whole-cell models~\cite{karr2012whole}, can be constructed by cross-referencing annotations across multiple databases to identify gene-protein-reaction (GPR) associations. 
Resources, such as the Integrated Microbial Genomes database'~\cite{markowitz2011img} and the SEED~\cite{overbeek2013seed}, maintain annotated genomes that enable automated generation of such models. 
Ultimately, more ambitious genome engineering projects, which move beyond refactoring and recoding into more complex changes of organism function, will likely need to make modeling a critical portion of the design process and will need to implement function not just at the sequence level but by composing separately characterized genetic parts and devices.
Much fundamental research still needs to be conducted, however, before such approaches will become practical at the gigabase genomic scale.

Additional constraints on design come from the methods used to modify genomes.
This necessitates adoption of consistent policies to resolve design conflicts, as well as to achieve the desired design objectives. Examples of design policies include the removal of elements, such as restriction sites, separation of overlapping features, replacement of codons, or optimization of nucleotide content for synthesis. In the Sc 2.0 project, design policies were semi-formalized through group consensus, but in practice were implemented by a combination of customized software tools or manual sequence editing.
Such policies could be enhanced by encoding synthesis expertise in the form of rules-based ontologies~\cite{venkatachalam1993knowledge, abrantes2017rule}. The BOOST tool developed by JGI provides a prototype for such a language~\cite{oberortner2016streamlining}.
Synthesis providers could also use such methods to encode their synthesis constraints, enabling automatic pre-qualification of designs as readily compatible with a providers' synthesis constraints.

The business interfaces of DNA synthesis providers could also improve.
Current tools typically operate at the level of individual fragments, often ordered by a human-centric web interface or, less frequently, APIs, with \emph{ad hoc} sequence redesign based on feedback from the tool on synthesis constraints. 
This makes it difficult for the designer to asses the overall impact of synthesis provider interactions in the context of the larger design, 
meaning that changes made to optimizing synthesis can inadvertently disrupt vital regulatory regions or sequences required for subsequent assembly, cloning, or sequence verification. 
Particular sequence features, such as repeats, secondary structure, and high GC content, can cause synthesis failures or problems assembling a DNA construct, often resulting in need for substantial redesign, project delays, and escalating costs.
DNA synthesis providers may also have constraints specific to their platforms.
Large-scale construction would be better supported by more standardized interfaces that would allow designers and synthesis providers to share a complete construct design in a common data format, 
including genomic context, annotation with desired functional biological features, and assembly plans (e.g., Gibson junctions). 
This would then allow a more collaborative interaction, in which the synthesis service can optimize for synthesis, or at least recommend solutions to the user, while remaining within the constraints of the original design and tracking the relation between suggested changes and the original target sequence. This also has an additional benefit for biosecurity, as the intent behind synthesis orders becomes more transparent.
The SBOL format is well-suited for supporting such annotations, though GenBank and GFF could also potentially be extended to encode this information.

\subsection{Building Engineered Genomes}
Technology and protocols to build engineered genomes specified in the design process are advancing rapidly. 
Still, the current approach of assembling DNA from smaller pieces is likely to hold for the foreseeable future up to the gigabase scale.
Building a gigabase genome will likely require synthesizing shorter DNA sequences, assembling these into larger constructs, and realizing the final genome that passes to the testing portion of the engineering workflow. 
Discussion of scalable protocols for building gigabase-scale genomes may be found in~\cite{boeke2016genome} and~\cite{TechnicalWorkingGroupInRevision}; here, we focus on the integration challenges of workflows with such protocols.

Depending on the specific host and intended function of the engineered organism, there are typically numerous alternative approaches and protocols for DNA synthesis, assembly, and delivery. These must result not only in the designed DNA sequence; synthesis, assembly, and delivery must proceed in a context that is ultimately compatible with and results in the stable expression of that genome within the host organism with the desired phenotype. This is no small task. Currently, there is an unmet need for guidance on best-practices for measuring, tracking, and sharing information regarding the engineered genome, as it is built step-by-step.

Building an engineered genome, then, involves manipulating DNA many times, which offers ample opportunities for reduced yield, breakage, error, and other sources of uncertainty in achieving the designed DNA sequence. Most DNA synthesis begins with shorter DNA fragments ordered from a service provider. 
Biases and other sources of uncertainty in the delivered sequence associated with a synthesis platform and method are not typically communicated in a useful fashion along with the delivered product. Rather, the DNA is often accompanied by a certificate asserting sequence verification. Sequencing by the recipient at this early step to verify and validate the received DNA is typically not done, due to cost and time constraints. 
However, the synthesized DNA may occasionally behave in biomolecular reactions in ways that are unexpected based on the requested sequence, such as during polymerase chain reactions with specific primers during amplification. This can be due to the presence of a secondary population of DNA with a slightly different sequence, which may not be apparent from sequencing (e.g., appearing as a second band after amplification in a gel electrophoresis separation). 
Large-scale operations should develop more comprehensive and quantitative approaches to measuring, tracking, and communicating information pertaining to quality control.

Protocols and commercial kits to assemble shorter DNA fragments into larger constructs often involve amplification, handling, purification, transformation, or other storage and delivery steps that can increase uncertainty regarding the quality and quantity of the DNA. Cell-based workflows may then incorporate growth and selection of colonies of cells that express the target constructs, thereby rejecting non-functional or unassembled DNA sequences that may have passed through the workflow up to this point. 
Still, depending on the approach, the resulting assembled DNA may include added sequences that are not biologically active, as in the case for some methods using restriction enzymes, or scars, such as occur may occur with Golden Gate Assembly~\cite{engler2009golden} or MoClo~\cite{weber2011modular}. Gibson Assembly~\cite{gibson2009enzymatic} is scarless, but the yield and specific results may depend on the secondary structure of the DNA fragments. 
Thus, information to track through the workflow should include the assembly method, sequences required for assembly, and their location along the DNA molecule, such as landing pads, sequences for compatibility with ``helper'' strains (typically \textit{E. coli} or yeast) used during construction, and DNA secondary structure. 
Larger DNA constructs are likely to be assembled using cell-based methods such as homologous recombination in yeast~\cite{gibson2008one}, and cell-based platforms that support eventual delivery to a variety of final host organisms and cell types would be desirable. 
Regardless of the specific approach, however, assembly must proceed in a context that supports the packaging, delivery, and eventual expression of the DNA in the target host, 
which may include imbuing the DNA with the correct epigenetic modifications for expression to achieve the desired phenotype. 

All this information must be expressed in the inputs provided from the design stage, and while FASTA or GenBank may be sufficient for a simple sequence, more sophisticated representations, such as GFF or SBOL~\cite{roehner2016sharing}, are needed for hierarchical construction.
SBOL can also convey functional information, allowing more manufacturing flexibility in the precise sequence chosen for construction, as well as full details of assembly plans and records~\cite{cox2018synthetic}, such as JGI manages with its BOOST software~\cite{oberortner2016streamlining}. 
Sequence modifications may be expressed in an enhanced sequence encoding language, such as BpForms~\cite{lang2019bpforms}.
Sample inventories of partially assemblies maintained as in vitro constructs or within strains may be exchanged with databases already designed and used for such purposes, such as ICE~\cite{ham2012design} and SynBioHub~\cite{mclaughlin2018synbiohub}.

The sequence of the assembled DNA is not typically verified at intermediate steps in the build process. However, good practice dictates that the DNA should be further cloned into a cell of interest, and colonies picked, sometimes at random, and the DNA sequenced to ensure assembly according to the intended design. Rather than sequence the entire DNA assembly, often only the specific region of interest is verified. Throughout the DNA assembly process, an abstracted description of the process would allow for machine reasoning over the combination of assembly methods and verifications to achieve scale. 
Sequence verification results are typically produced in FASTQ, which is generally sufficient for smaller constructs.
To operate on large-scale genomes, however, it may be important to assemble the information into more comprehensive descriptions of a genome and its variations using representations such as GVF~\cite{reese2010standard} or SBOL.

Suitable options for the delivery of large, assembled DNA constructs and whole genomes are generally lacking. The yield of existing processes, such as electrical and chemical transformation or genome transplantation, could be improved significantly to increase their utility, and a broader range of approaches should be developed for use with any organism and cell type. 
This may also require identifying new cell-free environments or cell-based chassis for assembling and manipulating DNA that also have compatibility with genome packaging and delivery systems into host organisms. Similarly to DNA assembly, delivery protocols and their associated information regarding number of biological and technical replicate experiments, methods, measurements, and so on, should be available in a machine-readable format. This should include information regarding the host cell, such as its genotype, which is often not fully verified. It is unclear the extent to which the background cellular context in which the assembled DNA is inserted affects its expression and the ultimate phenotype of that cell.

The adoption of best practices from industrial biomanufacturing settings could provide a path forward toward integrating appropriate measurements, process controls, and information handling for large collaborations involved in engineering entire genomes. Appropriate implementation of laboratory information management systems (LIMS) into research settings could go a long way toward organizing, archiving, tracking, and sharing information within and across the individual laboratories of a larger collaboration. 
Advancing the use of automation to support the build step of the genome engineering workflow requires evaluation of which steps may reduce costs and speed results, the availability of automated methods, ways to effectively share those methods and adapt them across platforms and manufacturers, and straightforward ways to integrate automated steps into a larger automated workflow, as well as integrate that workflow with machine learning. Issues concerning best-practices for cybersecurity, cyberbiosecurity, and biosecurity should also be considered and implemented. 
Much the same arguments apply for the testing discussed in the next section.

\subsection{Testing the function of engineered genomes}

Once the final DNA constructs have been integrated to produce an engineered genome in the desired biological context, both the sequence and expression of that genome should be verified to test for the intended phenotype. 
Collaborating organizations should agree on specific measurements, along with control and calibration measurements, to ensure that results can be compared and used across the participant laboratories towards achieving the final goal. 
The challenges of sequence verification of the engineered genome will be similar to those described above for intermediate DNA constructs during the build process.
In addition, to assess strain fitness and other phenotypic information, any number of omics and phenotypic measurements could be made, and measurements that return meaningful information to inform and advance the overall workflow should be prioritized. 
It will be necessary to decide at which scales of biological organization, such as single cell or population, and time, from milliseconds to days or years, the expression of the final DNA construct should be assayed to ensure the desired phenotype for the intended purpose. 
Increasingly, time-dependent attributes of cell processes and phenotype, as related to both gene expression internal to the cell and variable environmental conditions external to the cell, are expected to impact engineering decisions, including at the gigabase scale.

Details of particular assays are beyond the scope of this discussion, but it will likely be useful to develop standards and measurement assurance targeted for testing engineered genomes, for example, to help identify predictive relationships between genotype and phenotype or determine contributions of biological stochasticity and measurement uncertainty to overall variability in a measured trait. 
Standard protocols, reference cell lines, and the use of experimental design are examples of tools available to increase the rigor and confidence in conclusions drawn from testing. 
The development of methods for absolute quantitation would complement common relative measurements and facilitate calibration for comparability across measurement techniques and measurands along the central dogma. Calibration of biological assays will also aid in comparing results within a single laboratory over time and across different laboratories. 
Recent studies, for example, for fluorescence~\cite{wang2017standardization,beal2018quantification},  absorbance~\cite{stevenson2016general}, and RNAseq~\cite{lee2016external} measurements, demonstrate the possibility of realizing scalable and cost-effective comparability in biological measurements.
Best practices from biomanufacturing may be of use in informing all of these decisions.

DNA constructs are often evaluated for their associated growth phenotypes to determine the nature and extent of unexpected consequences for cell function and fitness due to the revised genome sequence. This is complicated by the need for accepted definitions of fitness, along with methods to measure and quantify fitness. Often, determining fitness effects also involve measurements of metabolic burden. Guidance on how to choose the best measure of fitness for a specific purpose or to speed the workflow would be helpful. Engineered cell lines should also be evaluated for robustness to changes in the environmental context the cells are likely to experience during typical use in the intended application, as we as stability over relevant timescales to evolution or adaptation. 

For a large collaboration, it will also be critical to establish shared representations and practices for metadata, process controls, and calibration.
Both the target fitness specifications and the protocols to evaluate them will need to be described in sufficient detail to enable automation-assisted integration and comparison of data, metadata, process controls, and calibration across laboratories.
These practices will help to ensure that testing includes measurements compatible with learning through modeling and simulation. Although some measurements will serve both test and learn portions of the engineering workflow, such as high-throughput metabolomics, some are likely to be different for testing the assembled DNA function and representing that function in a model.
In general, however, we expect that it will be valuable to have sufficient metadata to track the data, conditions, and samples all the way back to the designs, in a machine-readable manner to allow operations at scale.
Appropriate LIMS tooling and curation assistance software (e.g.,~\cite{wolstencroft2011rightfield}) will be vital for enabling such metadata to be created consistently, correctly, and in a timely fashion, by limiting the required input from human investigators.

\subsection{Learning systematically from test results}

Given our current limited knowledge of biology and current limited capabilities to design biology, learning the functions that genomes encode and learning how to design genomes will be critical to gigabase-scale genome engineering. 
This includes learning to predict the behavior of synthetic genomes, such as emergent interactions between new metabolic byproducts and existing machinery; learning fundamental biology, such as the functions of small non-coding RNA; learning design heuristics, such as preferentially using parts that only have one function; and learning better methods and processes, such as how to share information among teams of engineers. Developing the capabilities to predict and design the expression of genomes will likely become increasingly essential as our capability to construct synthetic organisms matures.

Design-build-test-learn workflows are an effective way to learn fundamental biology, design principles, and improved processes. The ideal workflow starts by using a predictive computational model, such as a whole-cell model \cite{karr2012whole} or a whole-organism model \cite{sarma2018openworm}, to design a genome; proceeds by constructing, booting up, and testing this genome; and concludes by systematically learning from behavioral failures of the genome by using the data generated from testing the genome to improve the model, our biological knowledge, and the methods used to design, construct, and boot up the genome. One way to systematically learn better models is to identify the minimum set of changes that must be made to the model to align its predictions with observed behaviors of the synthetic genome. 
To realize these models, we should develop new experimental methods for better characterizing the relationship between genotype and phenotype, new tools discovering and aggregating the data needed for modeling (building on foundations such as the workflow model introduced in SBOL 2.2~\cite{cox2018synthetic}, and ontology resources such as OBO~\cite{smith2007obo} and the Experimental Factor Ontology~\cite{malone2010modeling}), new formalisms for modeling and simulating the combinatorial complexity of biology and the multiple scales between genomes and organismal behavior, new methods for high-performance simulation of large models, new tools verifying models, new frameworks for collaboratively building models, and new representations for the semantic meaning and provenance of models \cite{goldberg2018emerging, szigeti2018blueprint}. 
To automatically learn these models from test data, we should develop new repositories of models of individual biological parts that can be composed into models \cite{cooling2010standard}; new methods for generating variants of models that explain new observations by incorporating models of additional parts, alternative kinetic laws, or alternative parameter values; and develop new model selection techniques for non-linear multiscale models \cite{kirk2013model}. 

Until we have comprehensive predictive models, engineers will likely rely on ad hoc combinations of predictive models of parts of organisms, data-driven models, and heuristic design rules. BioModels \cite{glont2017biomodels}, the NeuroML database \cite{crook2014model}, Open Source Brain \cite{gleeson2019open}, and the Physiome Model Repository \cite{yu2011physiome} already contain hundreds of models of parts of organisms that can be used to help design organisms. For example, metabolic engineers often use constraint-based models to design microbial factories \cite{kim2008metabolic}. Numerous data-driven models of parts of organisms can also already be used to help design synthetic organisms. For example, PSORTb \cite{yu2010psortb} can help bioengineers design signal sequences to traffic proteins to specific compartments. Several design heuristics are also already commonly used. 
For example, bioengineers often optimize the GC content of parts to minimize their metabolic load on their host \cite{villalobos2006gene}.

To close the design-build-test-learn cycle, the community should describe these models and heuristics with standard formats and share these models through public repositories. CellML \cite{cuellar2015cellml}, NeuroML \cite{gleeson2011development}, and Systems Biology Markup Language \cite{hucka2003systems} (SBML) formats can already be used to describe several types of biochemical and physiological models, and the BioModels, NeuroML, Open Source Brain, and Physiome repositories can already be used to share models. 
Similarly, Docker containers and the DockerHub repository~\cite{dockerhub} can already be used to share data-driven models. New formats for describing multiscale models should be developed to help bioengineers use these models to design genomes. In particular, these formats should support more comprehensive and more detailed models. In addition, these formats should facilitate collaboration by concretely capturing the semantic meaning of each model component and capturing the provenance of each model element, including the data sources, assumptions, and design decisions that motivated each element. Finally, to help bioengineers reliably design genomes, these formats should also capture the predictive performance of models, including the behaviors, biological mechanisms, and environments which they can and cannot capture.

\subsection{Coordinating complex workflows}

Engineering gigabase genomes will require coordinating numerous heterogeneous tasks into clear, cohesive, reproducible workflows \cite{myers2017standard, goni2016implementation}. Several technologies have already been developed to coordinate experimental protocols, computational workflows, and other complex tasks. Laboratory automation systems enable investigators to design and execute complex experimental protocols that involve numerous individual tasks, reagents, equipment, and personnel \cite{sadowski2016harnessing}. To execute protocols more reproducibly than possible with manual approaches, these systems capture the details of each task, such as the amount of each reagent, the order of tasks, and the results of each protocol. These systems can also be integrated with LIMS \cite{prasad2012trends} to help track workflows and reagent stocks. 
Individual tasks could be described with languages such as Autoprotocol~\cite{miles2018achieving}. 
Although lab automation has not yet been widely adopted, some researchers are already using lab automation systems, such as Aquarium \cite{keller2019aquarium} and Antha \cite{synthace2019antha}, to engineer yeast \cite{yang2019synthetic}. As these systems improve and as their interfaces are standardized and integrated, the community should adopt these tools for specifying build and test protocols. This would help large numbers of researchers execute standardized protocols collaboratively.

On the informational side, computational workflow engines aim to enable the clear specification and reproducible execution of complex computational workflows that involve multiple steps and require multiple software programs and computing environments. 
Like lab automation systems, workflow engines aim to capture the details of each task, the relationships among the tasks, and the dependencies of the tasks so that workflows can be shared among investigators and reproducibly executed. Several workflow engines such as Cromwell \cite{broad2019workflow}, Galaxy \cite{goecks2010galaxy}, NextFlow \cite{di2017nextflow}, and Toil \cite{vivian2017toil} have been developed for genomics. In addition, the Common Workflow Language \cite{amstutz2016common} (CWL) enables workflows to be exchanged among several of the most popular engines. 
Furthermore, the Dockstore \cite{o2017dockstore} and MyExperiment \cite{goble2010myexperiment} repositories enable researchers to publish workflows. These tools should be adapted for gigabase engineering workflows. To make genome engineering workflows easy to understand, we recommend that the community develop a graphical tool for designing genome engineering workflows and an ontology for annotating the semantic meaning of workflow tasks. To make genome engineering workflows easy to reproduce and reuse, we recommend that the community include CWL files in COMBINE archives~\cite{bergmann2014combine}, develop REST or other programmatic interfaces for all of the databases involved in genome engineering workflows, containerize~\cite{soltesz2007container} all of the computational tools needed for genome engineering, and deposit these containers to a registry such as DockerHub \cite{dockerhub}.

Issue tracking systems, such as GitHub issues \cite{githubissues}, can help teams coordinate the execution of complex tasks which are difficult to describe and automate. For example, software development teams frequently use issue tracking systems to coordinate the development of features of software. Issue tracking systems could be readily adapted to coordinate complex tasks involved in designing genomes and learning from tests of genome designs that require expert analysis. Furthermore, the PROV ontology~\cite{missier2013w3c} could be adopted to capture the provenance of these tasks, and to link information throughout the design-build-test-learn cycles, as is now possible in SBOL~\cite{cox2018synthetic}.

\subsection{Sharing data through common repositories}
Once links are established across different portions of a workflow, 
unified access to information in databases for various institutions and stages of the workflow
can be accomplished using standard federation methods and any of the various mature open tools for database management systems (DBMS).

To support collaborative sharing, genome-engineering consortia should adopt principles of FAIR (findable, accessible, interoperable, reproducible) data management. 
FAIR principles put specific emphasis on enhancing the ability of machines to automatically find and use the data, in addition to supporting its reuse by individual people. Some software supporting such principles for biological data sharing already exists, such as FAIRDOMHub~\cite{wolstencroft2016fairdomhub} and EDD~\cite{morrell2017experiment}, and may be adopted for this purpose.

Ultimately, the community should incorporate every aspect of the design-build-test-learn cycle into integrated, reproducible workflows. Such integrated workflows would help teams scale to gigabase genomes, develop extensible workflows that can be adapted for similar engineering projects, and conduct genome engineering transparently and reproducibly. Achieving this will require integrating lab automation systems for systematically executing build and test tasks, workflow engines for systematically executing design and learn tasks, and potentially issue tracking systems for managing more sophisticated tasks that require expert input.

\subsection{Laws and Contracts}

Few resources are available that can be adopted for coordinating the legal and contractual interactions of a large-scale genome-engineering project. 
However, some precedents could serve as the foundation for such resources.
In the realm of copyright law, the Open Source Initiative (OSI) and other software organizations have systematized licensing to provide compatible families of licenses, and these are used widely to promote collaboration and composability of software.
These licences can be classified into four tiers: public domain, permissive (e.g., BSD), copy-left (e.g., GPL) and proprietary. 
Similarly, the Creative Commons family of licenses provide a simple framework for media and other content sharing that has been widely adopted and is commonly incorporated into tooling.
With just a single badge in such an established system of licensing, a user or a machine can tell quickly if an object can be reused, if its reuse is prohibited, or if more complicated negotiation or determination is required. 

For dealing with physical biological materials, the first standardized materials transfer agreement was the NIH's Uniform Biological Materials Transfer Agreement (UBMTA), released in 1995.
The Addgene public DNA repository uses this agreement, which is widely supported by universities and enables many legal transactions to be automated, although not for commercial uses.
Broader and more compatible MTA systems have been developed by the Science Commons project~\cite{nguyen2007science}, and more recently by the OpenMTA~\cite{kahl2018opening}.
No publicly available system yet supports automation for management of protections for proprietary genetic materials and reagents, however, and there are also significant considerations to be managed regarding compliance with local regulatory and legal systems, particularly when materials cross international borders.

Developing a machine-compatible system of protections for scientific data and materials would enable transfers with automated systems and reduce institutional friction.
In order for such a system to work, the genome-engineering community should define tiered levels of IP protections that are simultaneously intelligible for the common user, legal experts, and computer systems. 
Enabling effective use of such systems with automation-assisted workflows will also require recording provenance information about which inputs are involved in the production of results, using mechanisms such as the PROV ontology~\cite{missier2013w3c}.
Finally, large-scale data sharing will also need to assist researchers and engineers in managing the level of exposure of information, whether due to issues of privacy, safety, publication priority, or other similar concerns.
Again, no current systems exist, but a basis for developing them may be found in the cross-domain information sharing protocols that have been developed in other domains \cite{chandersekaran2008cross, sun2009cross}.

\section{Recommendations and Outlook} \label{recommendations}

As discussed above, scaling up to gigabase genomes presents a wide range of immediate and long-term challenges (Table~\ref{t:stoplight}).
These cluster into four general areas, each with a different set of needs and paths for development:
\begin{itemize}
\item {\bf Representing and exchanging designs, plans, data, metadata, and knowledge:} 
Managing information for gigabase genomic design requires addressing many challenges regarding scale, representation, and standards. 
Relatively mature technologies exist to address most individual needs, as well as to assist with the integration of workflows.
The practical implementation of effective workflows will require significant investment in building infrastructure and tools that adopt these technologies, including domain-specific extensions and refinements.

\item {\bf Sharing and integrating data quality and experimental measurements:} 
Sharing and integrating information arising from measurements of biological material poses significant challenges. 
It remains unclear what information would be advantageous to share, given the difficulty of obtaining and interpreting measurements of biological systems and the expense and unfavorable scaling of data curation.
However, effective integration depends on associating measurement data with well-curated knowledge and metadata in compatible representations.
A number of potential solutions exist for each of these, but significant investment will be needed to investigate how the state of the art can be extended to satisfactorily address these needs.

\item {\bf Integration of modeling and design at the gigabase genomic scale:} 
Considerable challenges surround efforts to develop a deeper understanding of the relationship between genotype and phenotype, 
regarding both the interpretation of experimental data and the application of that data to create and validate models, which may be applied in computer-assisted design.
Long-term investment in fundamental research is needed, and the suite of biological systems of varying complexity, from cell-free systems to minimal and synthetic cells to natural living systems, may offer suitable experimental platforms to match genes with specific functions.

\item {\bf Technical support for Ethical, Legal, and Societal Implications (ELSI) and Intellectual Property (IP) at scale:} 
At the gigabase scale, computer-assisted workflows will be necessary to manage contracts, intellectual property, materials transfer, and other legal and societal interactions.
Such workflows will need to be developed by interdisciplinary teams involving experts in law, ELSI issues, software engineering, and knowledge representation.
Moreover, it will be critical to address these issues early.
\end{itemize}

In short, engineering gigabase-scale genomes presents significant challenges that will require coordinated investment to overcome.
Because many other areas of bioscience face similar challenges, solutions to these challenges will likely also benefit the broader bioscience  community.

Importantly, the challenges of scale, integration, and lack of knowledge faced in genomic engineering are not fundamentally different in nature than those that have been overcome previously in other engineering ventures, such as aerospace engineering and microchip design, which required organizing humans and sharing information across many institutions over time. 
Thus, we expect to be able to adapt solutions from these other fields for genome engineering.

Investment in capabilities for genomic engineering workflows is critical to move from a world in which genome engineering is a heroic effort to one in which genome engineering is routine, safe, and reliable.
Investment in workflows for genomic engineering will support and enable a vast number of projects, including many not yet conceived, as was the case for reading the human genome. As workflow technologies improve, we anticipate that the trends of Figure~\ref{fig:scale}B will eventually reverse, enabling high-fidelity whole-genome engineering at a modest cost.

\section{Competing Interests}
The authors declare no competing interests.

\section{Acknowledgements and Disclaimers}
This work was supported in part by NIH awards P41-EB023912 and R35-GM119771 and by NSF awards 1548123 and 1522074. 
The views, opinions, and/or findings expressed are those of the author(s) and should not be interpreted as representing the official views or policies of these funding agencies or the U.S. Government.
This document does not contain technology or technical data controlled under either U.S. International Traffic in Arms Regulation or U.S. Export Administration Regulations. 

Certain commercial equipment, instruments, or materials are identified to adequately specify experimental procedures. Such identification neither implies recommendation nor endorsement by the National Institute of Standards and Technology nor that the equipment, instruments, or materials identified are necessarily the best for the purpose.

\section{Author Contributions}
B.A.B., J.B., J.R.K., and E.A.S equally contributed to the article.

\section{References}
\printbibliography[heading=none]

\end{document}